\documentclass[aps,preprint]{revtex4}%
\usepackage{amsfonts}
\usepackage{amsmath}
\usepackage{amssymb}
\usepackage{graphicx}%
\begin{document}
\preprint{HEP/123-qed}
\title[Fractal vortex structure]{Fractal vortex structure in the lattice model of superconductors}
\author{S.A. Ktitorov}
\affiliation{A.F. Ioffe Physico-Technical Institute, the Russian Academy of Sciences,
Polytechnicheskaja str. 26, St. Petersburg 194021, Russia and Department of
Physics, St. Petersburg. State Electroengineering University, Prof. Popov str.
5, St. Petersburg 197376, Russia.}
\pacs{PACS number}

\begin{abstract}
The problem of the vortex structure in the lattice model of superconductors
has been reduced to the nonlinear map problem characteristic for the fractal theory.

\end{abstract}
\maketitle

%\volumeyear{year}

%\author{Second Author}

%\author{Third Author}
%\affiliation{Other Institution}

%\volumenumber{number}
%\issuenumber{number}
%\eid{identifier}
%\date[Date text]{date}
%\received[Received text]{date}

%\revised[Revised text]{date}

%\accepted[Accepted text]{date}

%\published[Published text]{date}

%\startpage{101}
%\endpage{102}
%\tableofcontents

\section{Introduction}

The lattice magnetic translation symmetry of superconductors near the upper
critical magnetic field was shown to be crucially important for the critical
thermodynamic properties \cite{1}, \cite{2}, \cite{3}. Attepts to construct
the incommensurate vortex structure were undertaken in \cite{4}. The small
parameter $\beta=\Phi/\Phi_{0}$ (where $\Phi=Ha^{2}$ and $\Phi_{0}=\frac{\hbar
c}{2e}$ are respectively the magnetic flux through the lattice plaquette
$a^{2}$ and the London quantum of magnetic flux) was essentially used there
that predetermined very weak dependence of the sample free energy on the
vortex configuration. Here I formulate this problem in a different way, so
that smallness of $\beta$ will be not assumed.

\section{Basic equation}

Confining myself here by the mean field approximation I consider the
two-dimensional model. Apart from neglecting of fluctuations this means that
an entanglement of vortices will not be accounted for in this paper. The free
energy functional for the lattice model reads:%
\begin{align}
F  &  =\sum_{n,n^{\prime};m,m^{\prime}}J_{nm,n^{\prime},m^{\prime}}\exp\left[
i\frac{2e}{\hbar c}\int d\mathbf{l\cdot A}\right]  \overline{\phi_{n^{\prime
}m^{\prime}}}\phi_{nm}+\label{free}\\
&  \sum_{nm}\left[  \tau\overline{\phi_{nm}}\phi_{nm}+g\left(  \overline
{\phi_{nm}}\phi_{nm}\right)  ^{2}\right]  ,\nonumber
\end{align}
where $J$ is the tunneling integral, the lattice site coordinates can be
written as $x=ma,y=na$, $\overline{\phi_{nm}}$ stands for the complex
conjugated order parameter, $\mathbf{A}$ is the vector potential, integration
of it is carried out along the straight line connecting the sites $nm$ and
$n^{\prime}m^{\prime},$ $\tau=\alpha\frac{T-T_{c}}{T_{c}}$. Differentiating
eq. (\ref{free}) with respect to $\overline{\phi_{nm}}$ we obtain the lattice
Ginzburg-Landau (LGL) equation for a superconductor at a strong magnetic
field:%
\begin{equation}
\sum_{n^{\prime}m^{\prime}}J_{nm,n^{\prime},m^{\prime}}\exp\left[  i\frac
{2e}{\hbar c}\int d\mathbf{l\cdot A}\right]  \phi_{n^{\prime}m^{\prime}}%
+\tau\phi_{nm}+g\left\vert \phi_{nm}\right\vert ^{2}\phi_{nm}=0 \label{LGL}%
\end{equation}
Choosing the Landau gauge $\mathbf{A=e}_{y}Hx$ and considering the simple
square lattice case within the tight-binding approximation (with $J$ standing
for the nearest-neighbor tunneling integral) we can write eq. (\ref{LGL}) in
the form%
\begin{align}
&  J\left\{  \phi_{m+1,n}+\phi_{m-1,n}+\phi_{m,n+1}\exp\left[  -i2\pi\alpha
m\right]  +\phi_{m,n-1}\exp\left[  i2\pi\alpha m\right]  \right\}
+\nonumber\\
\tau\phi_{nm}+g\left\vert \phi_{nm}\right\vert ^{2}\phi_{nm}  &  =0.
\label{nhe1}%
\end{align}
Eq. (\ref{nhe1}) can be transformed to the simpler form. Notice that eq.
(\ref{nhe1}) is a nonlinear equation; therefore, the Fourier transforming is
not particularly useful since it will convert this equation into an
integral-difference one. However, the specific form of nonlinearity
characteristic for the Ginzburg-Landau equation allows us to make a
substitution that is usually done in the theory of the linear Harper equation
\cite{5}:%
\begin{equation}
\phi_{mn}=u_{m}\exp\left(  i\kappa m\right)  . \label{sub}%
\end{equation}
Then eq. (\ref{nhe1}) takes the form%
\begin{equation}
u_{m+1}+u_{m-1}+2\cos\left(  2\pi m\beta-\kappa\right)  u_{m}+\tau
u_{m}+g\left\vert u_{m}\right\vert ^{2}u_{m}=0. \label{nhe2}%
\end{equation}
Eq. \ref{nhe2} can be named the nonlinear Harper equation (NHE) or the
Harper-Ginzburg-Landau (HGL) equation. The rest of the paper will be devoted
to a discussion of possible consequences of this equation.

\section{Analysis of the nonlinear Harper equation}

Eq. \ref{nhe2} is very peculiar one. It contains two mechanisms, either of
which can lead to forming of irregular self-similar structures that have
fractal properties. If we shall proceed in the spirit of the Abricosov vortex
structure theory, we have firstly to linearize the equation.%

\begin{equation}
u_{m+1}+u_{m-1}+2\cos\left(  2\pi m\beta-\kappa\right)  u_{m}+\tau u_{m}=0,
\label{lhgl}%
\end{equation}
This equation differs from the linear Harper equation \cite{5} in the same way
as the Ginzburg-Landau one for a superconductor at the magnetic field differs
from the Schroedinger equation for a charged particle at the magnetic field.
Therefore, the line $H_{c2}$ is determined by the condition $\tau=\epsilon
_{l}$, where $\epsilon_{l}$ is the Harper multiband spectrum lower edge. The
dimensional crossover takes place near the lowest band bottom at $\beta$
rational \cite{1}. The flux structure is completely pinned by the crystal in
this case. A subsequent decrease of temperature or magnetic field, i.e. an
increase of $|\tau|$, leads to appearance of fractal structures. The Harper
operator spectrum is known to form a Cantor set at irrational $\beta$ and,
therefore, the dimensional crossover is absent in this case. Scaling
properties of this operator have been studied using the renormalization group
technique \cite{6, 7}. Exact solutions with a use of the Bethe Ansatz and
quantum groups can be found in \cite{8}. Multifractal properties of the Harper
operator eigenfunctions and eigenvalues are discussed basing on numeric
calculations and on the Bethe Ansatz in the work by Wiegmann with co-authors
\cite{9}. Interesting results have been also obtained in \cite{10}.
Unfortunately, the Bethe Ansatz rigorous approach is not extended to the
nonlinear case; it is a challenge now for the mathematical physics community.
An approximate solution of the linearized Harper equation can be presented as
a chaotic distribution of the Wannier-type functions, which can be considered
as nucleation centres of the superconducting state. The order parameter
amplitude can be found perturbatively in this case. On the other hand, Eq.
(\ref{nhe2}) can be written in the form of the two-dimensional nonlinear map
on the complex field:
\begin{equation}
\left(
\begin{array}
[c]{c}%
u_{m+1}\\
u_{m}%
\end{array}
\right)  =\left(
\begin{array}
[c]{cc}%
-\tau-g\left\vert u_{m}\right\vert ^{2}-2\cos(2\pi m\beta-\kappa) & -1\\
1 & 0
\end{array}
\right)  \left(
\begin{array}
[c]{c}%
u_{m}\\
u_{m-1}%
\end{array}
\right)  . \label{map}%
\end{equation}

\section{Conclusion}

The main inference from the work is that the fractal flux structure is
controlled by two-dimensional maps on the complex field enumerated by the
whole real numbers. This means that the two-dimensional flux lattice in the
crystal lattice can be completely specified establishing the nucleation centre
distribution along one say leal axis. In opposite to the continuum model case,
this distribution is not periodic in general. The following distribution can
be taken as a first approximation{\large
\begin{align}
\phi(x,y) &  =C\exp\left(  -\frac{x^{2}+y^{2}}{2l_{\,H}^{\,2}}\right)
\times\label{abric}\\
&  \sum_{n=-\infty}^{\infty}\exp\left[  -\pi\left(  n+u_{n}\right)  ^{2}%
+\frac{\sqrt{2\pi}i\left(  x+iy\right)  }{l_{H}}\left(  n+u_{n}\right)
\right]  ,\nonumber
\end{align}
}{\normalsize \bigskip} where $u_{n}$ are determined by some proper discrete
map taking fractal effects into account. Notice that this sum is reduced at
the condition $u_{m}=0$ into the elliptic theta function so that this formula
can be considered as its natural generalization on special number fields
similarly to introduction of the basis functions, for instance. Along with the
discussed above map, one can try the Chirikov standard map, which is
equivalent to the Frenkel-Kontorova model that is widely used for description
of the incommensurability effects:
\begin{equation}
\left\{
\begin{array}
[c]{l}%
I_{n+1}=I_{n}-\frac{2\pi V_{0}}{\lambda b}\sin\left(  2\pi y_{n}/b\right)  \\
y_{n+1}=y_{n}+a+I_{n+1})
\end{array}
\right.  \label{map3}%
\end{equation}
or, if the interaction between vortices must be taken into account, the
generalized standard map
\begin{subequations}
\begin{equation}
\left\{
\begin{array}
[c]{l}%
I_{n+1}=I_{n}-\frac{2\pi V_{0}}{\lambda b}\sin\left(  2\pi y_{n}/b\right)  \\
y_{n+1}=y_{n}+a-\frac{1}{\beta}\ln(1-\beta I_{n+1})
\end{array}
\right.  .\label{map4}%
\end{equation}

\begin{acknowledgements}
I would like to thank E. Kudinov, Yu. Kuzmin and B. Shalaev for discussions.
The work was supported by the Russian Foundation for Basic Research, grant No 02-02-17667.
\end{acknowledgements}

\end{subequations}

\end{document}